\documentclass[twocolumn,prl,superscriptaddress]{revtex4} 

\usepackage[latin3]{inputenc}
\usepackage{graphicx}
\usepackage{verbatim}
\usepackage[english]{babel}
\usepackage{amssymb}

\begin{document}
\author{Antonio Scialdone}
\affiliation{Dipartimento di Scienze Fisiche, Universit\`a di Napoli ``Federico II'', Italy}
\affiliation{INFN, Napoli, Italy}
\author{Mario Nicodemi\footnote{Corresponding author. E-mail: mario.nicodemi$@$na.infn.it}}
\affiliation{Dipartimento di Scienze Fisiche, Universit\`a di Napoli ``Federico II'', Italy}
\affiliation{INFN, Napoli, Italy}

\title{Diffusion-based DNA target colocalization by thermodynamic mechanisms}

\begin{abstract}
In eukaryotic cell nuclei, a variety of DNA interactions with nuclear elements occur, which, in combination with intra- and inter- chromosomal cross-talks, shape a functional 3D architecture. In some cases they are organised by active, i.e. actin/myosin, motors. More often, however, they have been related to passive diffusion mechanisms. Yet, the crucial questions on how DNA loci recognise their target and are reliably shuttled to their destination by Brownian diffusion are still open. Here, we complement the current experimental scenario by considering a physics model, in which the interaction between distant loci is mediated by diffusing bridging molecules. We show that, in such a system, the mechanism underlying target recognition and colocalization is a thermodynamic switch-like process (a phase transition) that only occurs if the concentration and affinity of binding molecules is above a threshold, or else stable contacts are not possible. We also briefly discuss the kinetics of this ``passive-shuttling'' process, as produced by random diffusion of DNA loci and their binders, and derive predictions based on the effects of genomic modifications and deletions.\\
\end{abstract}

\maketitle

\section*{Introduction}

In the nucleus of eukaryotic cells, the spatial organization of chromosomes has a functional role in genome regulation  \cite{deLaat2003,Dekker2008, Fraser2007,Takizawa2008,Kumaran2008,Meaburn2007,Misteli2001,Misteli2008}: DNA loci, for a correct activity, must occupy specific, but dynamically changing, positions with respect to other DNA sequences or nuclear elements. A diverse number of interactions exist but the mechanisms whereby distant loci recognize each other and come together in complex space-time patterns are still largely unknown. Examples are found of loci that undergo directed motion via active, i.e. actin/myosin-dependent, processes \cite{Chuang2006,Dundr2007,Dekker2008,Kumaran2008,kolomeisky2007,spudich2001,Lonard2008}. 
However, most examples of cross-talks appear to be independent of active motors. Therefore, passive diffusion has been proposed as a major, energetically inexpensive, mechanism \cite{Misteli2001,deLaat2003,Dekker2008}. Brownian mobility induces stochastic collisions of loci, which, in turn, establish functional associations, e.g. via bridging molecules. Such a scenario, however, raises fundamental questions \cite{deLaat2007,Meaburn2007, NicodemiTerritories}. How are these random encounters coordinated in space and time? Are they probable? Are they reliable for functional purposes? How are they regulated?
\bigskip

Complex regulatory inter-chromosomal contacts occur, for instance, in the $\beta$-globin $T_{H}2$ Hox clusters \cite{Lanzuolo2007, Palstra2003}. Another striking example is observed during X chromosome inactivation (XCI) in female mammalian cells. At the onset of XCI, the X inactivation centre ($Xic$) regions on the two Xs come in close apposition to regulate expression of the $Xist$ gene \cite{Xu2007, Xu2006}. The $Xic$ interaction is mediated by the $Tsix/Xite$ (and $Xpr$) \cite{Augui2007} locus and relies on an RNA-protein bridge including CTCF, a zinc-finger protein having a cluster of a few dozen binding sites at the locus \cite{Xu2007}. Once the different fates of the active and inactive X chromosome have been determined, they are then targeted to different nuclear positions: the active X to the nuclear envelope and the inactive one, by $Xist$, to the nucleolus for maintenance of its silenced state \cite{Zhang2007}.
Many other cases are known. The loop architecture of the major histocompatibility complex class I (MHC-I) locus on human chromosome 6 \cite{Kumar2007} is mediated, for instance, by a set of specific molecules. Here, chromatin loops are organised by SATB1 and PML proteins, and PML-associated nuclear bodies, which tether clustered DNA binding sites to the nuclear matrix. The number and position of these anchoring regions depend on the relative abundance of SATB1 and PML protein \cite{Kumar2007}. For example, whereas Jurkat T cells show five chromatin loops within such a region, CHO cells, having a lower expression of SATB1, have six loops that also differ in position \cite{Galande2007, Kumar2007}. However, if the SATB1 concentration in CHO cells is matched with that of Jurkat T cells, a new loop organisation miming that of Jurkat T cells is found \cite{Kumar2007}.
Looping of specific remote loci is fundamental for the regulation of the $Kit$ gene in erythropoiesis (the production of red blood cells) \cite{Jing2008}. In immature erythroid cells, where $Kit$ is active, a distal $5'$ enhancer is shuttled to the $Kit$ gene promoter and bound by GATA2 proteins. Upon cell maturation, $Kit$ is repressed and the above conformation changed: GATA2 is displaced while GATA1 proteins and cofactors bring a downstream region to the promoter \cite{Jing2008}. In this case, the relative expression level of GATA proteins acts on the chromatin conformation and controls the switch of $Kit$ \cite{Jing2008}. Interestingly, clusters of binding sites are typically involved in most of the above examples \cite{deLaat2003, Fraser2007, Meaburn2007, Misteli2008, NicodemiTerritories}.
As mentioned earlier, the current question concerns the underlying organisational principles of such complex systems: how can Brownian random processes be finely regulated? How can such a variety of molecular elements be orchestrated? How do they recognise each other from a distance and get brought in apposition?
Here, we investigate a schematic physics model describing the interactions of a DNA locus (modelled as a polymer) and a nuclear target (e.g. nucleolus) mediated by a set of binding Brownian molecules. We show that target recognition and colocalization occurs via a switch-like thermodynamic mechanism - a phase transition - marked by specific thresholds in molecular binders concentration and affinity. Below these thresholds, diffusion is unable to produce colocalization; above these thresholds, despite the diffusive nature of motion, colocalization proceeds spontaneously at no energetic cost, with resources being provided by the thermal bath. Importantly, we show that binding energies and concentrations where the transition happens fall in the relevant biological range, whereas the ON-OFF character of the transition ensures the full reliability of the process. For this reason, this could be seen as a ``passive-shuttling process'', where the adjective ``passive'' should distinguish it from the form of shuttling produced by active motors (e.g. actin/myosin systems). Thus, our picture can explain how well-described cell strategies of upregulation of DNA binding proteins or chromatin chemical modifications can produce efficient and sharply regulated genomic architectural changes. The scenario we depict also has a close analogy with the known problem of polymer adsorption at a surface (see \cite{deGennes, deBell1993, Joanny1979, Semenov1996} and references therein). We describe the theoretical bases of the mechanism by a mean-field analytical approach, which we confirm by extensive Monte Carlo computer simulations. Finally, we briefly discuss the system kinetics.

\section{Materials and Methods}
\subsection*{The model}

We study two schematic models representing the situation where a DNA locus is shuttled towards a different nuclear target (e.g. nucleolus, nuclear membrane, matrix) or to another DNA sequence.
In the first case, the DNA sequence is represented, via a standard polymer physics model, as a floating random walk polymer of $n$ beads \cite{DoiEdwards}  
(fig.~\ref{fig2} upper panels). The polymer interacts with a concentration, $c$, of Brownian molecular factors (MFs) and can be bound at a number, $n_{0}$, of clustered binding sites (BSs) with chemical affinity $E$. In real examples, the number and location of binding sites depend on the specific locus considered. For definiteness, here we refer to the well-studied $Tsix/Xite$ locus of X colocalization and choose the number and chemical affinity of binding sites accordingly (see below). However, as known in polymer physics, our results are robust to parameter changes (see \cite{DoiEdwards} and below). In our model, a nuclear target is also included. It is schematically described as an impenetrable surface having a linearly arranged set of binding sites for the DNA binding molecules (Fig. \ref{fig2}, upper panels). For the sake of illustration, we assume that their number is also $n_{0}$ and their affinity $E$.
We use a simple lattice version of the random walk polymer model. This is well established in polymer physics and has the advantage to be simple enough to permit comparatively faster simulations with respect to off-lattice models. In this way, we can add further degrees of freedom into our system, which represent the binding molecules, without making computation unfeasible. In fact, molecules are dealt with as a statistical mechanics ``lattice gas'' interacting with the polymer chain \cite{Stanley}. We consider a cubic lattice of linear size $L_{x}=2L$, $L_{y}=L$ and $L_{z}=L$ (in units of $d_{0}$, the characteristic size of a bead on the polymer; see below), with periodic boundary conditions to reduce boundary effects \cite{Binder1997}. For the sake of simplicity, the DNA sequence is treated as a directed polymer \cite{DoiEdwards}, i.e., its tips are bound to move on the top and bottom surfaces of the system volume (Fig.~\ref{fig2}). It comprises $n=L$ beads, which randomly move under a ``non-breaking'' constraint: two proximal beads can sit only in the next or nearest next neighbouring lattice sites. A bond between an MF and a BS can be formed when they are on next neighbouring sites; MFs can have multiple bonds (such as with CTCF proteins). The use of directed polymers to represent DNA segments allows faster simulations without affecting the general properties of the colocalization mechanism we describe because they are produced by a general free-energy minimization mechanism, which does not depend on such details (see Results and Discussion sections). In the case of a non-directed polymer model, DNA would bind its target as well, but without a perfect alignment as in our model \cite{NicodemiTerritories} (Fig.~\ref{fig2}, upper panels). A strategy to attain a straight alignment anyway would be to consider a gradient of BSs along the polymer and its target. In real cells, the number and distribution of binding sites depend on the specific locus considered but, as shown in polymer physics \cite{DoiEdwards,Stanley}, our thermodynamic picture is robust.
To investigate the colocalization of two DNA sequences, we also consider a variant of a model where the nuclear scaffold is removed and a replica of the polymer is added \cite{NicodemiMeiosis, NicodemiPairing, ScialdonePairing}.
We explore these models by a statistical mechanics mean-field treatment and by Monte Carlo (MC) computer simulations. We try to use the available biological data to set the range of model parameters. Our models include only minimal ingredients and are very schematic, but they permit to derive a precise, quantitative picture of passive shuttling. Conversely, our scenario relies on a robust thermodynamic mechanism and its general aspects are thus not affected by the simplicity of the models. 

\subsection{DNA binding site number and chemical affinity}
Details on binding energies and DNA locations of binding sites are known in some examples (see \cite{BergJ2008a, Gerland2002, Lassig2007, Maerkl2007, Massie2008, Morozov2005} and references therein), but in most cases only qualitative information is currently available. 
For instance, in vitro measures exist \cite{Quitschke2000, Renda2007} for dissociation constants of CTCF proteins from DNA binding sites, which give binding energies around $E\sim20kT$, $k$ being the Boltzmann constant and $T$ the room temperature (for example, see  \cite{Zhao2009} on how to derive the binding energy from the dissociation constant). The precise value of in vivo binding energies depends on the specific DNA site considered and can be very hard to record, yet these in vitro measurements provide the typical energy range. It is experimentally well documented that DNA binding proteins, like those mentioned in the Introduction, have a number of target loci with chemical affinities in the weak biochemical energy range, $E\sim0-20kT$  \cite{BergJ2008a, Gerland2002, Lassig2007, Maerkl2007, Massie2008, Morozov2005}. This is the energy scale we consider here. Here, the BS number $n_{0}$ on the DNA, as well as on its target, is chosen to be $n_{0}=24$ (i.e., the order of magnitude of the known presence of CTCF sites in the $Tsix/Xite$ region on the X chromosome \cite{Donohoe2007}, but it is varied to describe the effects of BSs deletions, see fig.~\ref{fig5} inset). 

\subsection{Molecule concentration}
The order of magnitude of the concentration of molecular factors, $c$, can be roughly estimated and compared with the concentrations of proteins in real nuclei. In our model, the number of molecules per unit volume is $c/d_{0}^{3}$, where $d_{0}$ is the linear lattice spacing constant, which implies that the molar concentration is $\rho=c/(d_{0}^{3}N_{A})$, where $N_{A}$ is the Avogadro number. Under the assumption that a polymer bead represents a DNA segment of $\sim20$bp (i.e. of the order of magnitude of a CTCF binding site in $Tsix/Xite$ region) \cite{Donohoe2007, Xu2007}, we obtain the order of magnitude of the polymer bead size, $d_{0}\sim10nm$. By using such a value of $d_{0}$, typical concentrations of regulatory proteins such as $\rho\sim 10^{-3}-10^{-1} \mu mol/l$ 
(i.e., $\sim10^{3}-10^{5}$ molecules per nucleus) would correspond to volume concentrations in our model of $c\sim10^{-4}-10^{-2}$ percent. Such an estimate is approximate, but could guide the connection of our study to real biological situations.

\subsection{Monte Carlo simulations}
In our Monte Carlo (MC) simulations, we run up to $10^{9}$ MC steps per simulation and our averages are over up to $500$ runs. At each MC step, the algorithm tries to move, on average, all the particles of the system (molecules and polymer beads, in random order) according to a transition probability proportional to $e^{-\Delta H/kT}$ \cite{Binder1997}, where $\Delta H$ is the energy barrier of the move. Therefore, the binding/dissociation rate is given by the Arrhenius factor $r_{0}e^{-\Delta H/kT}$, where $r_{0}$ is the bare reaction rate. The MC time unit (a single lattice sweep) corresponds thus to a time $\tau_{0}=r_{0}^{-1}$ \cite{Binder1997}. In turn, $\tau_{0}$ is related to the polymer diffusion constant $D$ and to the lattice spacing constant $d_{0}$: 
$D=\left(\langle \Delta s^{2}\rangle d_{0}^{2}/4\tau_{0}\right)$, where $\langle \Delta s^{2}\rangle$ is the mean square displacement (expressed in units of $d_{0}$) of the polymer center-of-mass per unit MC time. We measure $\langle \Delta s^{2}\rangle$ and the value of $d_{0}$ can be estimated to be of the order of magnitude of a typical protein binding site, $\sim10 nm$ (see above). We impose that the diffusion coefficient $D$ of a free polymer (i.e. with $E=0$) in our lattice is of the order of magnitude of the measured diffusion constant of human DNA loci ($D=1 \mu m^{2}/hour$) \cite{Chubb2002}. As a result, an MC lattice sweep is found to correspond to $\tau_{0}\sim30 ms$ (falling well within the range of known biological kinetic constants \cite{Watson} ).
The above MC simulations produce an artificial dynamic and, in general, serious caution must be taken to interpret it as the real kinetics. However, in the current prevailing interpretation \cite{Binder1997}, in a system dominated by Brownian motions, an MC Metropolis dynamic is supposed to describe well the general long-term evolution of the system. Under that umbrella, we assume here that MC simulations could provide some insight into the system kinetics. We consider a lattice with $L=32$, i.e. with dimensions $L_{x}=2L=64$ and  $L_{y}=L_{z}=L=32$ in units of $d_{0}$. DNA segments have $n=32$ beads. We also performed simulations with different values of $L$ and $n$ (up to $L=128$ and $n=128$) and checked that our general results remained essentially unchanged. The conceptual support for using comparatively small self-avoiding walk (SAW) polymer chain sizes to extrapolate the behaviour of longer chains is grounded in statistical mechanics and relies on the system scaling properties \cite{Binder1997}. For instance, the transition energy $E^{*}$ has a comparatively simple behaviour with variations in $n$ \cite{NicodemiTerritories} and rapidly converges at a large $n$ to a finite value comparable with $E^{*}(n=32)$. Those remarks support the idea that our results are not an artefact of the specific length of the polymer.

\section*{Results}

\subsection*{Mean-Field theory} 

\begin{figure*}
\includegraphics[width=0.7\textwidth]{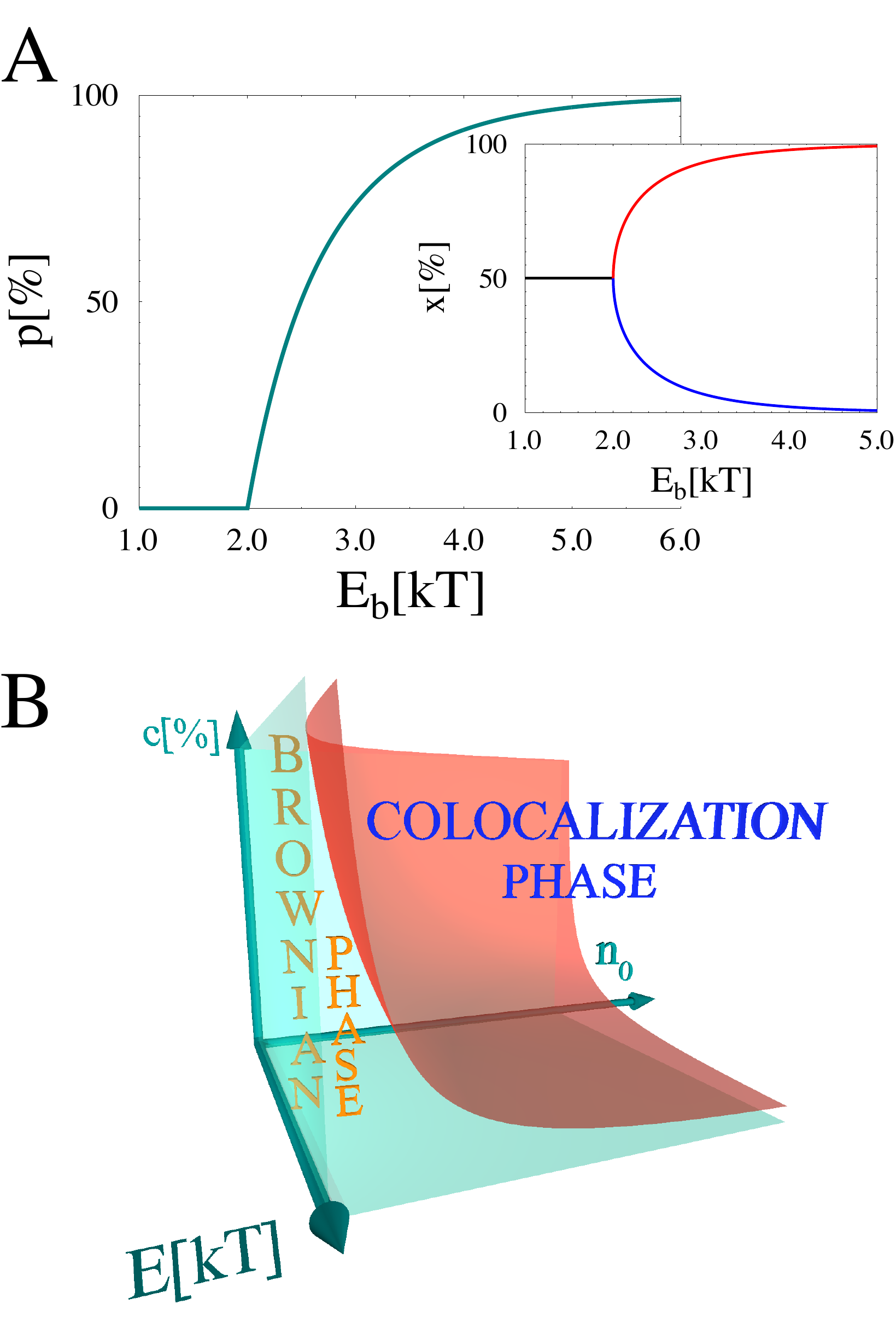}
\caption{\textbf{Thermodynamics 
of colocalization.}  The mean-field theory description of DNA-target colocalization. \textbf{Panel A} The polymer excess colocalization probability, $p$, is plotted as a function of the average binding energy density $E_{b}$. The \textbf{inset} shows the probability $x$, to find polymer 1 in the right half of the nucleus. The plots show how, at $E_{b}/kT=2$, a transition occurs between a phase where polymers are independently located in space $(p=0\%$, $x=50\%$) and a phase where they colocalize ($p>0\%$, $x\neq50\%$). \textbf{Panel B} The transition surface $cn_{0}E_{b}/kT=constant$ is depicted in the space of molecule concentration, $c$, binding energy, $E$, and number of binding sites, $n_{0}$.}\label{fig1}
\end{figure*}

To describe the concept behind passive shuttling and colocalization, we briefly discuss the statistical mechanics of the system at the level of a mean-field, coarse-grained approximation \cite{Stanley}. We refer to the polymer adsorption literature for more advanced theoretical approaches (\cite{deGennes, deBell1993, Joanny1979, Semenov1996} and references therein). For the sake of definiteness, we consider the case with two DNA polymers. We partition the nucleus into two halves and name $x$ the probability to find polymer $1$ in the right half and $y$ the probability to find polymer 2 in the left half. In a Ginzburg-Landau approach \cite{Landau, Stanley}, the system free-energy density can be written as a function of $x$ and $y$: $F\cong F(x,y)=H(x,y)-TS(x,y)$. The interaction energy density, $H$, can be expanded in powers of $x$ and $y$ to consider the first nontrivial terms: $H=-E_{b}\left[x(1-y)+y(1-x)\right]$.
The above quadratic form arises because a molecular bridge between the polymers can be formed only if they are in the same part of the nucleus. $E_{b}$ is the average binding energy density, which at low $c$ and $E$ is approximately the product of the density of available binding sites bound by a molecule, $cn_{0}$, multiplied by the total chemical affinity of a bridge, $2E$: $E_{b}(c, E, n_{0}) \propto 2Ecn_{0}$. In turn, the entropy, $S(x,y)$, in such a mean-field approach can be approximated as the sum of the entropies of the two non-interacting polymers, $S(x,y)=S(x)+S(y)$, where $S(x)$ has the standard expression $S(x)=-k[xln(x)+(1-x)ln(1-x)]$ \cite{Stanley}.
The equilibrium state of the system is obtained (in the thermodynamic limit) as the minimum of $F(x,y)$. The corresponding equations, $\partial_{x}F=\partial_{y}F=0$, always have a trivial solution $(x,y)=(1/2,1/2)$, representing the state where the polymers have independent and equal probabilities to be on the left or right side of the nucleus. However, if the bridging energy $E_{b}$ is larger than a critical value, $E_{b}^{*}=2kT$, the above solution turns into a saddle point and two new non-trivial minima arise where $x=1-y=1/2$ (Fig. \ref{fig1}A, inset). A second-order phase transition \cite{Stanley} occurs at $E_{b}^{*}$, with a consequent spontaneous symmetry breaking: the two minima, i.e. the thermodynamically favoured states, correspond to the colocalization of the polymers on the same side of the nucleus. The system order parameter is the polymer excess colocalization probability, $p$, i.e. the probability to find them in the same region minus the probability to be in different regions: $p=x(1-y)+y(1-x)-[xy+(1-x)(1-y)]$.
If $E_{b}<E_{b}^{*}=2kT$, polymers are independently located in the nucleus and $p=0$; above the critical point, they are more likely to be found together in the same area and $p>0$ (Fig. \ref{fig1}A). The critical energy value, $E_{b}^{*}$, corresponds to the point where the entropy loss owing to colocalization is compensated by the corresponding energy gain.
Close to the transition, $p$ has a power law behaviour, $p\sim[E_{b}-E_{b}^{*}]^{-\beta}$, with a mean-field exponent $\beta=1$ \cite{Stanley}. The phase where $p>0$ is the ``colocalization phase'', whereas for $p=0$, polymers move independently (the ``Brownian phase'').
The critical value $E_{b}^{*}=2kT$ can be written in terms of the model parameters  $(c, E, n_{0})$, providing the following expression for the transition surface (Fig.\ref{fig1}B): $cn_{0}E/kT=constant$.
The advantage of the above mean-field description is to illustrate the basic ideas of the scenario we propose. However, it is very schematic and in the following sections we discuss a detailed MC simulation of the model.

\subsection*{DNA-target colocalization}

\begin{figure*}
\includegraphics[width=0.8\textwidth]{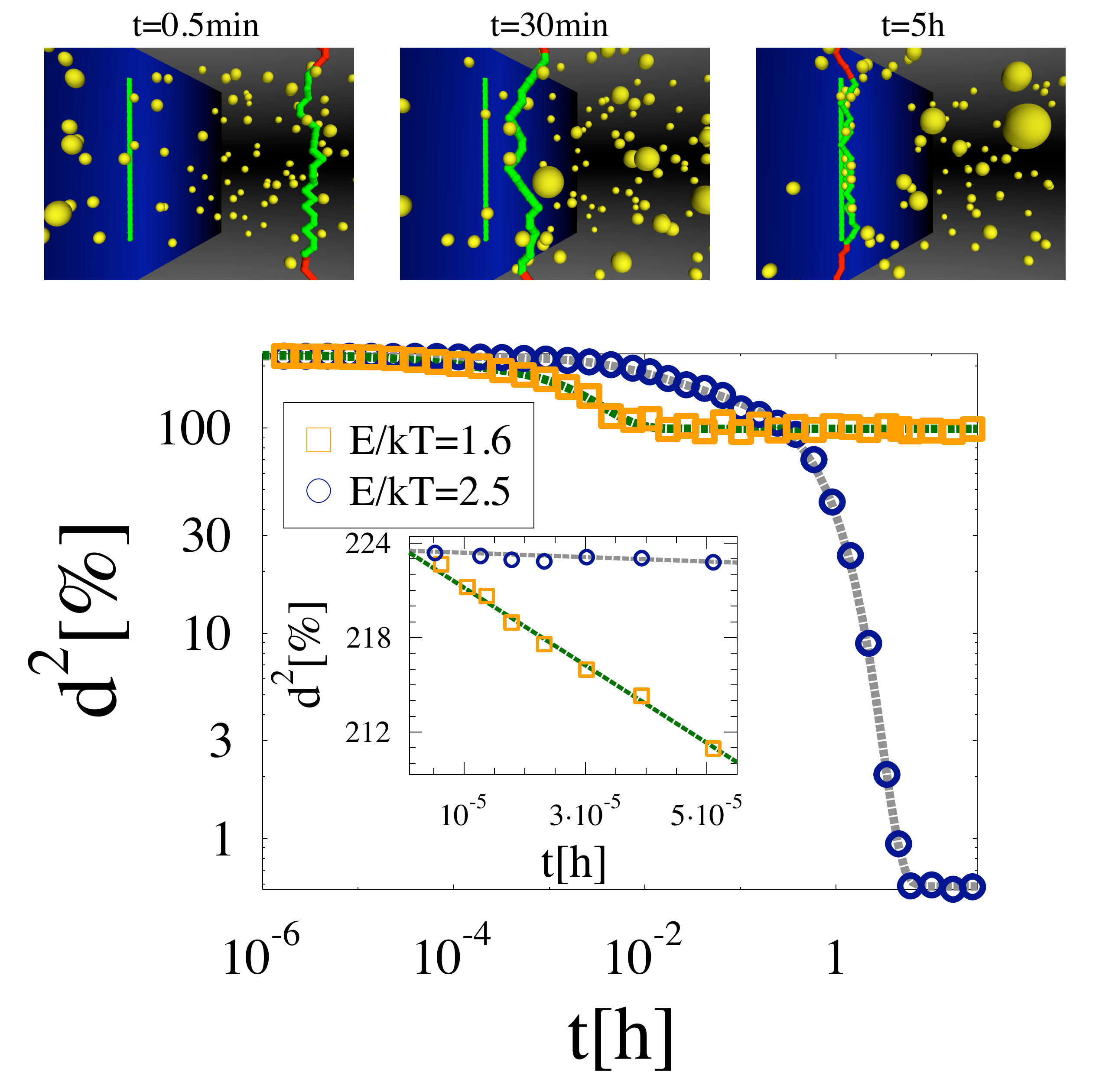}
\caption{\textbf{Kinetics of colocalization.} The normalized mean square distance, $d^{2}$, between the DNA and the nuclear target binding sites (BSs) is plotted as a function of time, $t$, for two values of their binding molecule chemical affinity, $E$ (here $c=0.2\%$ and $n_{0}=24$). Data are from Monte Carlo (MC) simulations. $d^{2}(t)$ has a linear diffusive behaviour at short $t$ (inset) and a long-time exponential approach to an equilibrium value. The latter corresponds to colocalization only if $E$ is above a threshold (see text and Fig.~\ref{fig4}). The upper panels show system configurations from MC simulations at three time periods for $E=2.5kT$ and provide a pictorial representation of our model: the DNA locus is modelled as a SAW polymer made by ``beads'' that have an affinity equal to $0$ or to $E$ (red beads and green beads, respectively) for Brownianly diffusing molecules (yellow beads). A cluster of binding sites is also present on the nuclear target (blue surface).}\label{fig2}
\end{figure*}

\textbf{The colocalization mechanisms - } 
We first consider the model describing the system made of the binding molecular factors, a polymer and a plane representing the surface of a nuclear target (Fig. \ref{fig2}, upper panels). Because the diffusing molecules can bind both the polymer and the plane, they can induce an effective attraction force between them via the formation of bridges. We illustrate such an effect by considering the mean square distance $d^{2}(t)$ between the binding sites (BSs) of the polymer sequence and the nuclear target BSs, as a function of time, $t$:
\begin{displaymath}
d^2(t)=\frac{1}{n_0}\sum_{z=1}^{n_0}\frac{\langle r^2(z,t) \rangle}{r^2_{rand}},
\end{displaymath}
where  $r^2(z,t)$ is the square distance between two BSs at height $z$ at time $t$, averaged over all the BSs and over different MC simulations (indicated by $\langle...\rangle$). We use 
as a normalization constant the mean square distance from target expected for a randomly diffusing polymer, $r^2_{rand}$. 
The system evolves according to the master equation simulated by the MC evolution. The DNA polymer is initially positioned at a distance $L$ from the target plane in a straight, vertical configuration (therefore, the starting value of the distance is $d^{2}(t=0)=L^{2}/r^{2}_{rand}\sim 2.5$) and a given concentration $c$ of MFs is randomly positioned in the volume. Fig.~\ref{fig2} shows the dynamics of $d^{2}(t)$ for two values of the interaction energy $E$ (here, $c=0.2\%$ and $n_{0}=24$). When the binding energy $E$ is small enough, say  $E=1.6kT$, the long-time value of $d^{2}(t)$ is equal to $100\%$, as expected for a randomly diffusing polymer. In principle, a bridge can be stochastically formed by an MF but no stable interactions are established, although a finite concentration of MFs is present. A drastic change in behaviour is observed, however, when the energy is raised to $E=2.5kT$ (fig. \ref{fig2}): now the long-time plateau of $d^{2}$ collapses to zero, signalling that a full colocalization has occurred. An effective attraction force is generated and the polymer spontaneously finds and stably binds its target (Fig. \ref{fig2}, upper panels).
The dynamics is characterised at short times by a Brownian diffusion regime where $d^{2}(t)$ is linear in $t$ (Fig. \ref{fig2}, inset) and the polymer randomly explores the space around. During that time, it enters into contact with its target. Afterwards, an exponential decay of $d^{2}(t)$ is observed to the equilibrium value and, for a large-enough $E$, the interaction is stabilised. A fit function for $d^{2}(t)$, which incorporates the initial linear and the later exponential regime, is:
\begin{equation}
d^2(t) = d^2(\infty) + \left[d^2(0)-d^2(\infty) + \frac{at}{1+bt}\right]\exp(-t/\tau) 
\end{equation}

where $d^{2}(\infty)$ is the thermodynamic equilibrium value of $d^{2}$ and $a$, $b$ and $\tau$ are fit parameters (depending on the value of $c, \, n_{0},\, E$). In particular, $\tau$ represents the average time needed to reach the asymptotic state, which increases as function of $c$ and $E$ as a consequence of the decrease in the DNA diffusion constant, as described in the next section.
Time scales can be affected by other complexities (such as chromatin entanglements, crowding, etc.) that we do not consider at the level of our schematic description here. However, it is interesting to note how, although only reasonable guess values are used for the system parameters (i.e. molecular concentration and/or affinity, number of binding sites, etc.; see model description), the values of $\tau$ predicted by MC simulations are compatible with the characteristic time scales of cellular processes ($\tau\sim 10-10^{2}$ minutes; Fig. \ref{fig2}) \cite{Watson}.\\

\begin{figure*}
\includegraphics[width=0.8\textwidth]{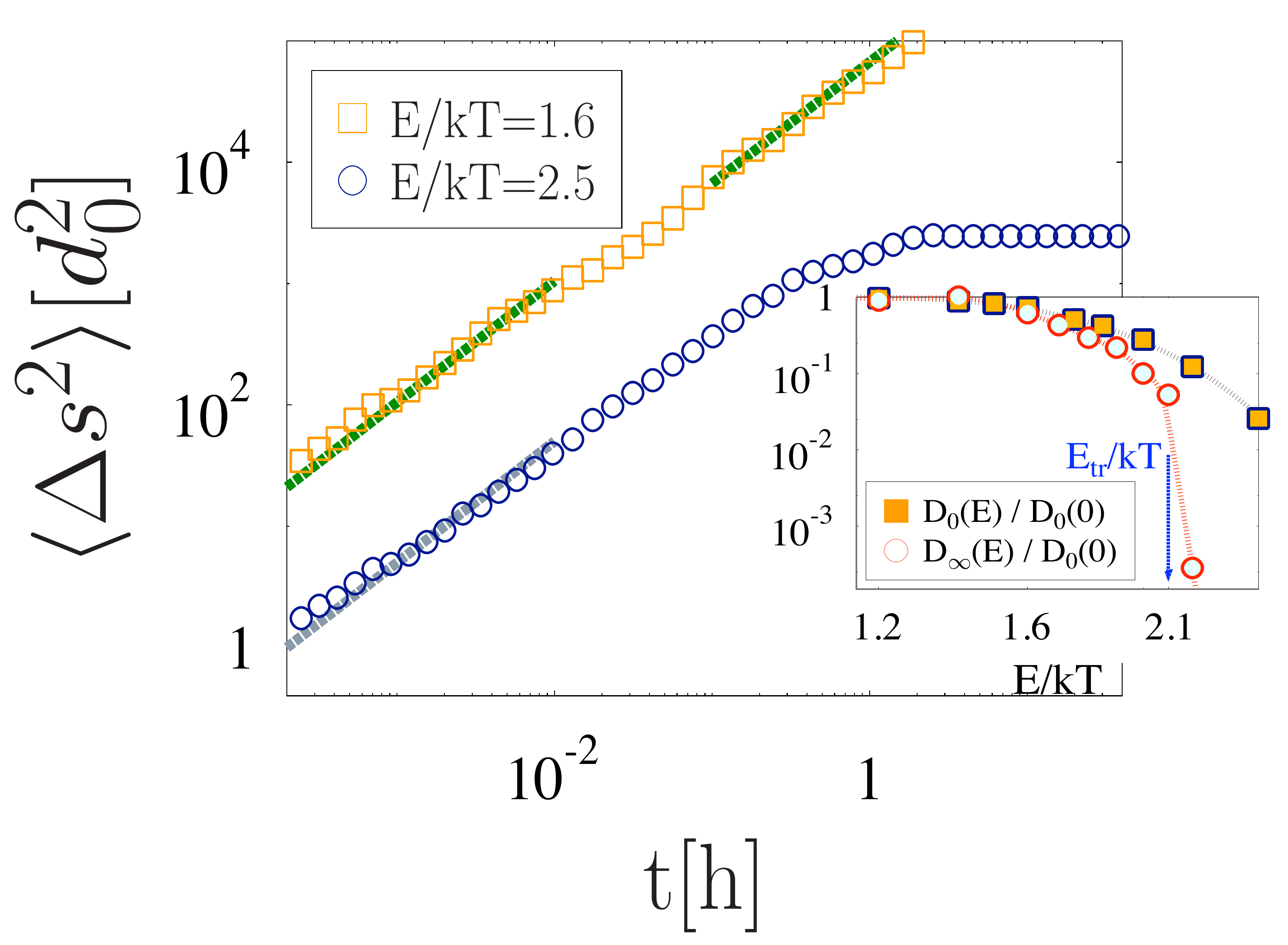}
\caption{
\textbf{DNA diffusion, recognition and colocalization to a nuclear target.} 
 The mean square displacement, $\langle\Delta s^{2}\rangle$, of the center-of-mass of the DNA polymer is plotted as function of time. While at $E/kT=1.6$ (squares) $\langle\Delta s^{2}\rangle$ is linear in $t$ at short as well as at long times, with the same diffusion constant ($D_{0}$ and $D_{\infty}$, inset), for $E/kT=2.5$ (circles) around $t=2$ hours, $\langle\Delta s^{2}\rangle (t)$ reaches a plateau, showing that a stable contact with the target is established and diffusion is interrupted. In the inset, the short- and long-time diffusion constants, $D_{0}$ and $D_{\infty}$, are shown as a function of $E$ (normalized by $D_{0}(E=0)$).}\label{fig3}
\end{figure*}

\textbf{Target search by diffusion - } 
The dynamics of colocalization is interesting in itself and still experimentally largely unexplored. As it is diffusive in nature, the mean square displacement from the initial position, $\langle\Delta s^{2}\rangle$, of the polymer center-of-mass is a main quantity describing the kinetics. Fig. \ref{fig3} shows $\langle\Delta s^{2}\rangle(t)$ for the same two values of the interaction energy, $E$, considered in the previous section. For $E/kT=1.6$ (i.e., when no stable colocalization is observed), $\langle\Delta s^{2} \rangle(t)$ has the typical Brownian linear behaviour with $t$, at short as well as at long times (Fig. \ref{fig3}); overall, the polymer motion is unaffected by the presence of binding molecules and nuclear scaffold. For $E/kT=2.5$, Brownian diffusion is instead only found at short $t$ while the polymer is searching for its target (Fig. \ref{fig3}, inset); at longer times, $\langle\Delta s^{2}\rangle(t)$ reaches a constant plateau, which signals that the polymer has become firmly bound to the scaffold BSs and cannot diffuse anymore (Fig. \ref{fig3}).\\
The inset in Fig. \ref{fig3} shows both the short- and long-time diffusion constants $D=\langle\Delta s^{2}\rangle(t)/4t$ (named, respectively, $D_{0}$ and $D_{\infty}$) as a function of the energy $E$. $D_{0}(E)$ has a smooth decreasing behaviour with $E$ because the larger the binding energy the higher the number of MF bound to the polymer. The long-time diffusion constant, $D_{\infty}(E)$, has a different behaviour. When E is small, $D_{\infty}(E)$ is very close to $D_{0}(E)$, showing that the polymer motion is diffusive at all times. However, above a transition point, $E_{tr}<2.1kT$, $D_{\infty}(E)$ collapses to zero as a result of the attachment of the DNA segment to the fixed scaffold, which stops further diffusion (Fig. \ref{fig3}, inset).

\begin{figure*}
\includegraphics[width=0.65\textwidth]{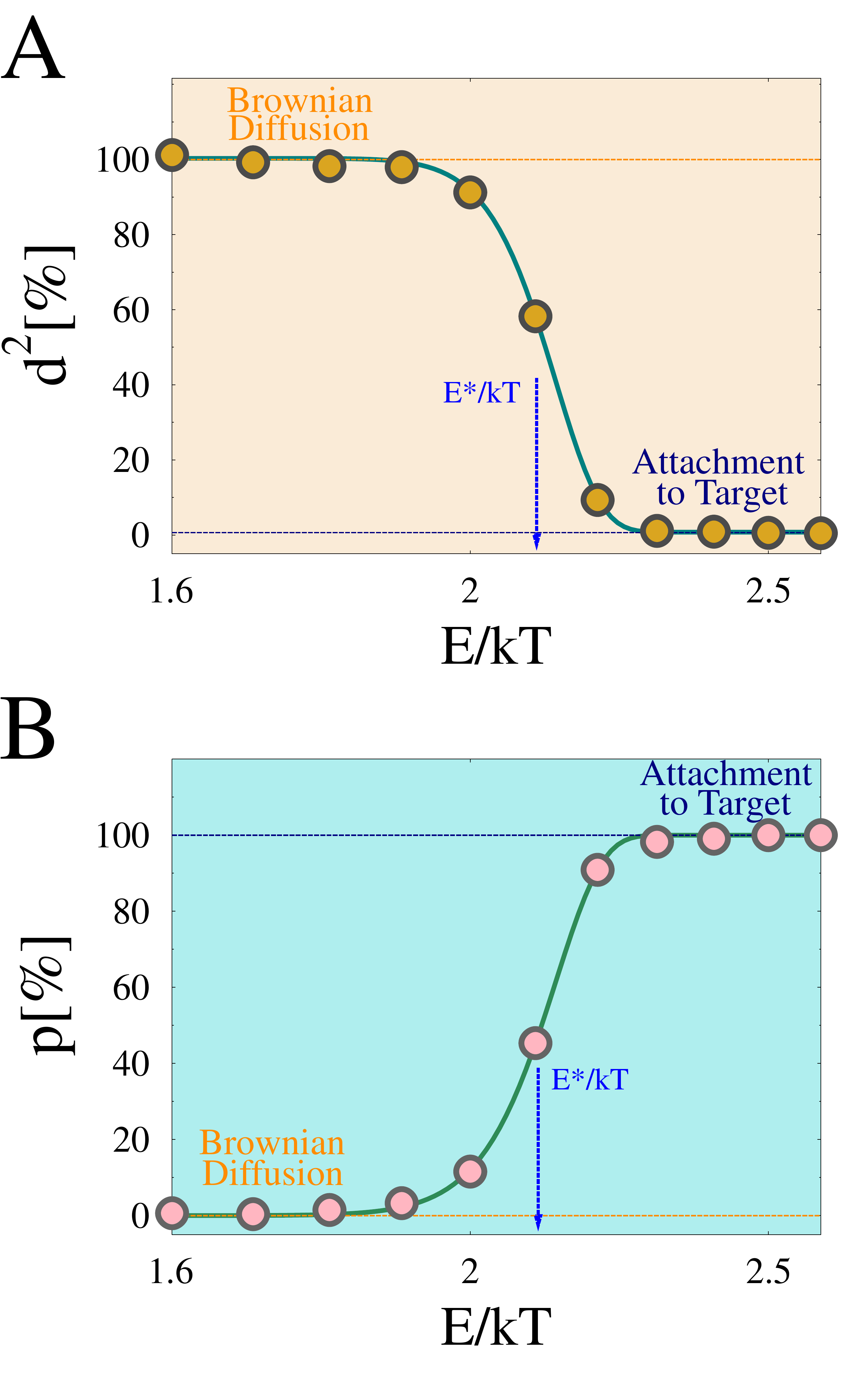}
\caption{ \textbf{The switch for DNA-target colocalization.} \textbf{Panel A} The equilibrium normalized square distance, $d^{2}$, between the DNA polymer and its nuclear target is shown as a function of $E/kT$, the binding molecule affinity. At small $E$, $d^{2}$ has a value corresponding to random diffusion ($d^{2}=100\%$); above a threshold $E\sim E^{*}=2.1kT$ (blue vertical arrow), a phase transition occurs and $d^{2}$ collapses to zero (blue horizontal line), indicating that DNA and target are colocalized. Molecule chemical affinity (or concentration, see Fig.~\ref{fig5}) acts as a switch. Here, $c=0.2\%$ and $n_{0}=24$. \textbf{Panel  B} The attachment probability, $p$, of DNA to target increases correspondingly from $0\%$ to $100\%$.
}\label{fig4}
\end{figure*}

\begin{figure*}
\hspace{-1.5cm}\includegraphics[width=0.9\textwidth]{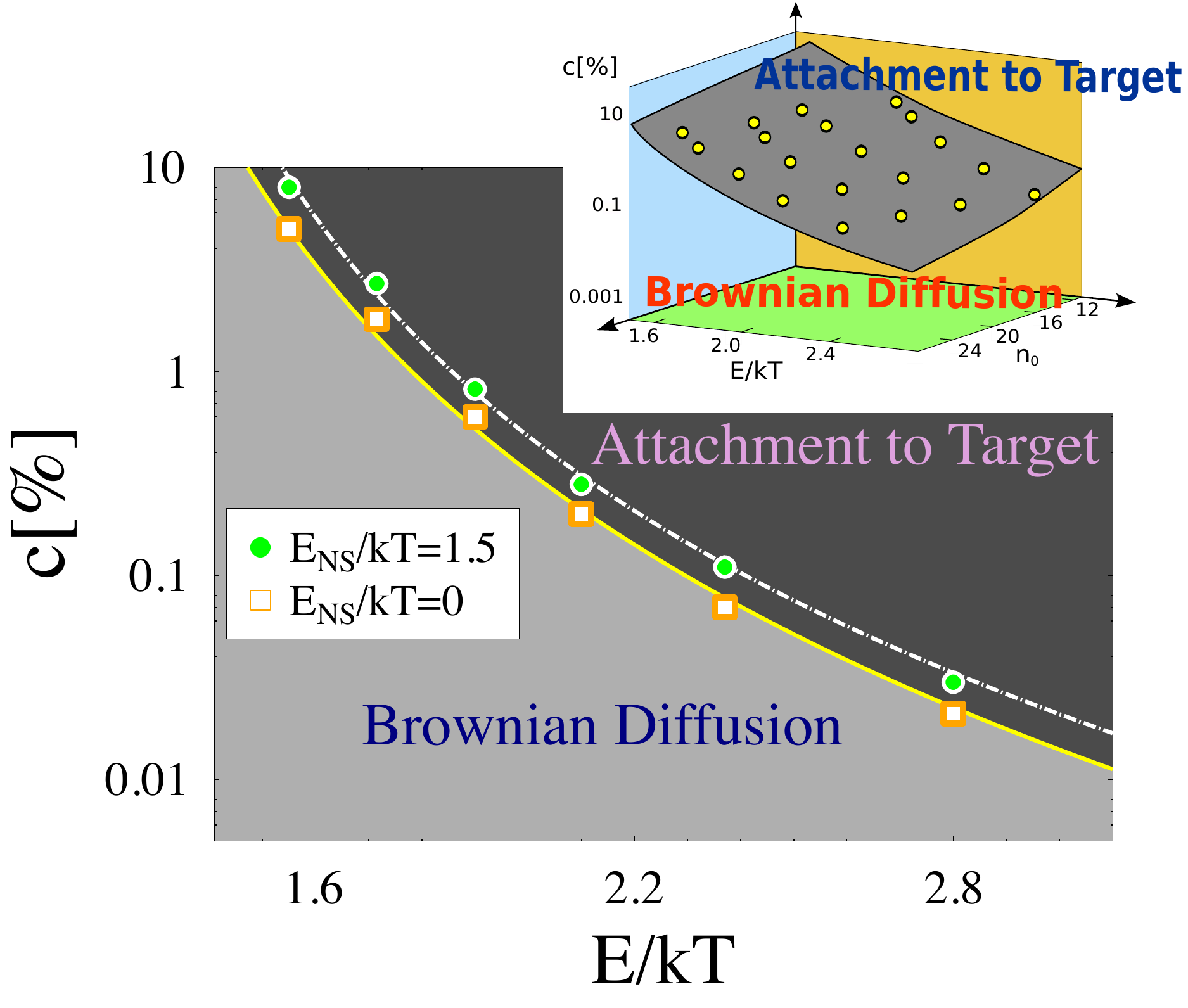}
\caption{\textbf{Colocalization-state diagrams.} The system phase diagram in the main panel shows the regions in the $(c,E)$ plane where DNA attachment to target and Brownian diffusion occur (here $n_{0}=24$), in the presence (green circles) or absence (orange squares) of non-specific binding sites with a low affinity for molecular binders ($E_{NS}=1.5kT$). In the inset, the full 3D phase diagram in the $(c,E,n_{0})$ space is shown. The transition surface (grey) is a power-law fit (see text).}\label{fig5}
\end{figure*}

\textbf{The switch for colocalization - } 
As much as the squared distance, $d^{2}$, we consider the probability, $p$, that the polymer is bound to its nuclear target, i.e. its mean distance from the scaffold BSs is less than $10\%$ of the lattice linear dimension $L$. The equilibrium values of $d^{2}$ and $p$ have a similar threshold behaviour as functions of $E$, as illustrated in Fig. \ref{fig4}. At small values of $E$, we find $d^{2}\simeq 100\%$ and $p\simeq 0$; conversely, for $E$ larger than a threshold $E^{*}\simeq 2.1kT$, $d^{2}\simeq 0$ and $p \simeq 100\%$, showing that stable binding to the target has occurred. The critical value $E^{*}$ is here defined by the criterion $p(E^{*})=50\%$ and is numerically consistent with the threshold, $E_{tr}$, found for $D_{\infty}(E)$. At approximately $E^{*}$, $d^{2}$ and $p$ have an intermediate value between that of the Brownian phase ($d^{2}\sim100\%$ and $p\sim0\%$) and that of the colocalization phase ($d^{2}\sim0\%$ and $p\sim100\%$); in this crossover regime (typical of phase transitions in finite systems) \cite{Stanley}, only partially stable bridges are built between polymer and nuclear target. Whereas the sharpness of the crossover region is known to increase in the thermodynamic limit \cite{Binder1997}, we found that it does not depend on the specific value of the system parameters (i.e. molecule concentration and affinity, and number of DNA target binding sites).
As predicted by mean-field theory, colocalization can also be triggered by an above-threshold concentration of MFs, $c$, and BS number $n_{0}$. This is illustrated by the phase diagram in $(c, E, n_{0})$ space (Fig.~\ref{fig5}, inset). The transition surface between the two phases has been obtained by a power-law fit of numerical data: $c(E-\tilde{E})^{\alpha}(n_{0}-\tilde{n}_{0})^{\beta}=constant$, with $\alpha=4.5$, $\beta=1.2$, $\tilde{E}=kT$ and $\tilde{n}_{0}=8$. From these data, we derive that, for typical concentrations of regulating proteins (i.e. $c\sim10^{-4}-10^{-2}$ percent, see above), the transition energies fall in the range $E\sim3-7kT$, which is well within the interval of typical DNA-protein affinities found in the literature ($E\sim0-20kT$).
Note that in a real population of cells, the fraction of colocalized sequences is expected to be smaller than in our in silico model for several reasons, such as the lack of full synchronization (DNA colocalization can be induced or released at different times in different cells, whereas our system is perfectly synchronized).

\begin{figure*}
\hspace{-1.5cm}\includegraphics[width=0.8\textwidth]{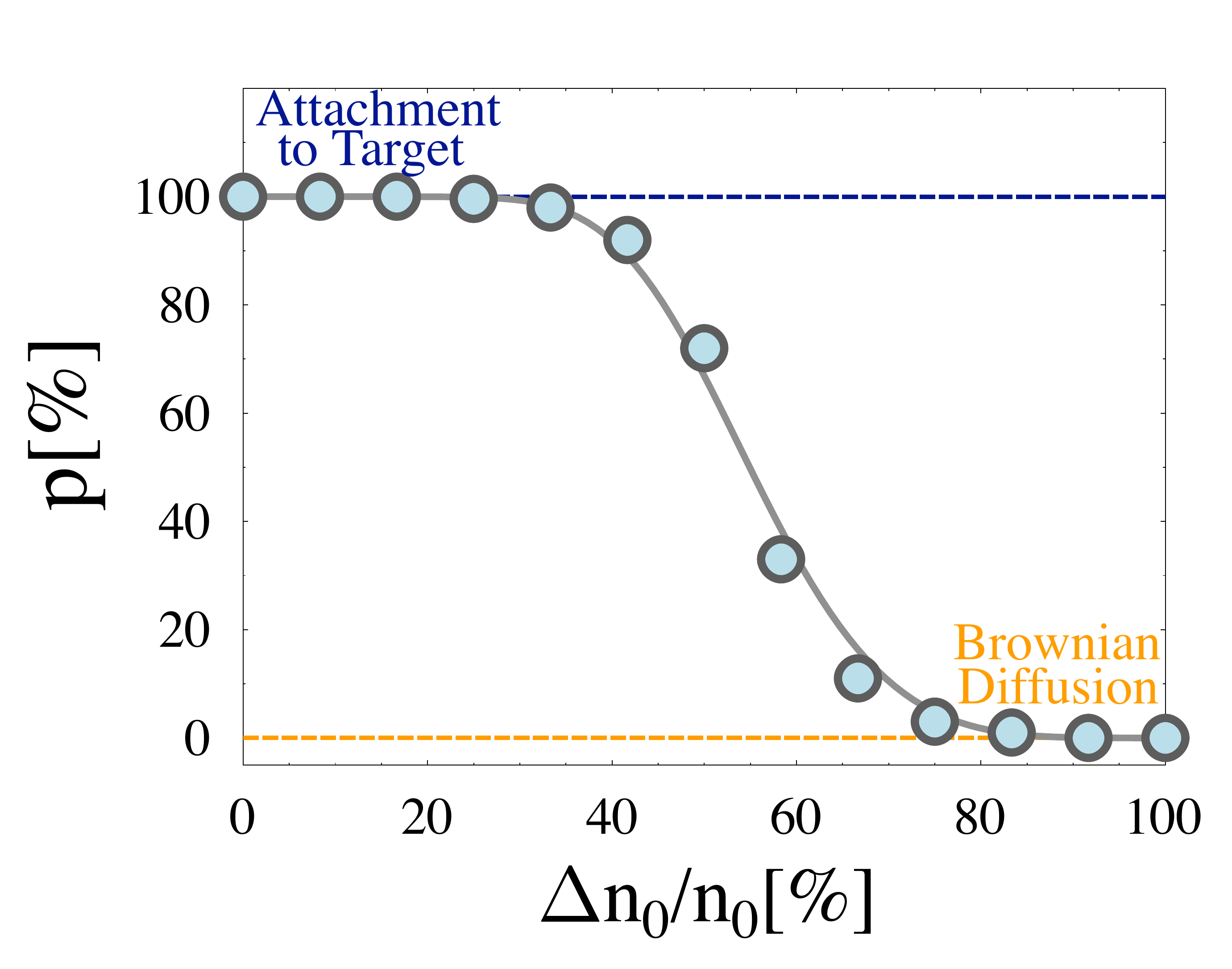}
\caption{
\textbf{Non-linear effects of deletions.} After deletion of a fraction, $\Delta n_{0}/n_{0}$, of DNA binding sites, the probability, $p$, of DNA-target colocalization is changed. $p$ has a non-linear behaviour with $\Delta n_{0}$: colocalization is only impaired by above-threshold deletions (here $E=2.5kT$, $c=0.2\%$ and $n_{0}=24$).}\label{fig6}
\end{figure*}

\textbf{Non linear effects of deletions - }
In our model, a variation in $n_{0}$ can describe the deletion of a fraction of DNA binding sites, and we consider now the case where the BSs on the polymer are reduced by an amount, $\Delta n_{0}$, with respect to the wild-type number $n0$. Deletions have a non-linear effect, characterized by a threshold behaviour. The equilibrium value of $p$ has a sigmoid shape, $\Delta n_{0}/n_{0}$, with a threshold of $\sim 50\%$ (Fig.~\ref{fig6}): short deletions (e.g. with $\Delta n_{0}/n_{0}<50\%$) do not result in a relevant reduction of $p$, whereas colocalization is lost as soon as $\Delta n_{0}/n_{0}$ gets larger than such a threshold. This ON/OFF behaviour stems from the nontrivial thermodynamic origin of the MF- mediated effective attraction between polymer and nuclear target. The threshold value $n_{0}$ is a decreasing function of $E$ and $c$, as seen in the phase diagram in the Fig.~\ref{fig5} inset. These results are predicting a nontrivial effect of deletions that could be tested experimentally. Similar results were also found for transgenic insertions and related ectopic association (data not shown).
In summary, we showed that binding MFs induce an effective attraction between the BSs on the DNA polymer and the other nuclear element, whereby the DNA segment is brought in close apposition with the target. The attraction, however, is only present if the MF concentration, the BS number and the MF-BS interaction energy are above a threshold value, otherwise the DNA segment randomly diffuses into the lattice (Fig. \ref{fig8}).

{\bf Role of non-specific binding sites - }
It is interesting to try to describe the effects on the colocalization mechanism of the presence of a number of non-specific binding sites on DNA and/or its target. The problem of how sequence- specific proteins can find their DNA sites on very large eukaryotic genomes is ancient (for example, see \cite{BergO1988,Lin1975,Hippel1986}). It has been proposed that the presence of non-specific binding sites allows a mixture of one-dimensional diffusion of bridging molecules along the DNA and three-dimensional diffusion in the surrounding medium, which could result in a more efficient search of the DNA target sites than a purely one- or three-dimensional diffusion \cite{BergO1981,Gerland2002,Slutsky2004,Winter1981a,Winter1981b}. Conversely, binding of molecules to these sites is expected to impair shuttling by the reduction of the effective concentration of diffusing molecular mediators. We tested the effect of the presence of non-specific sites in our schematic model: along with the clusters of specific sites previously included on the polymer and on its target, we inserted up to $4X10^{4}$  non-specific (i.e., low affinity) binding sites distributed on the target surface and within the polymer itself. We performed Monte Carlo simulations to find out the equilibrium status of the system as function of the molecular concentration $c$ and the specific binding energy $E$, with a fixed affinity for non-specific sites equal to $E_{NS}=1.5kT$.
Fig.~\ref{fig5} shows the changes in the phase diagram with respect to the case $E_{NS}=0$. Orange squares and green circles mark the transition points between the Brownian and the colocalization phase, respectively, for $E_{NS}=0$ (the case we dealt with previously) and $E_{NS}=1.5kT$. The plot reveals that the presence of non-specific binding sites moves the transition line upwards. This is due to a reduction in the effective concentration of molecules that are available to the specific sites, responsible for recognition and attachment to the target. This effect can be important and affects the location of the transition line even for comparatively small affinities (e.g. $E_{NS}=1.5kT$; Fig.~\ref{fig5}). However, the overall colocalization mechanism we discussed before is shown to be very robust.

\subsection*{Colocalization of DNA loci}
\begin{figure*}
\includegraphics[width=0.8\textwidth]{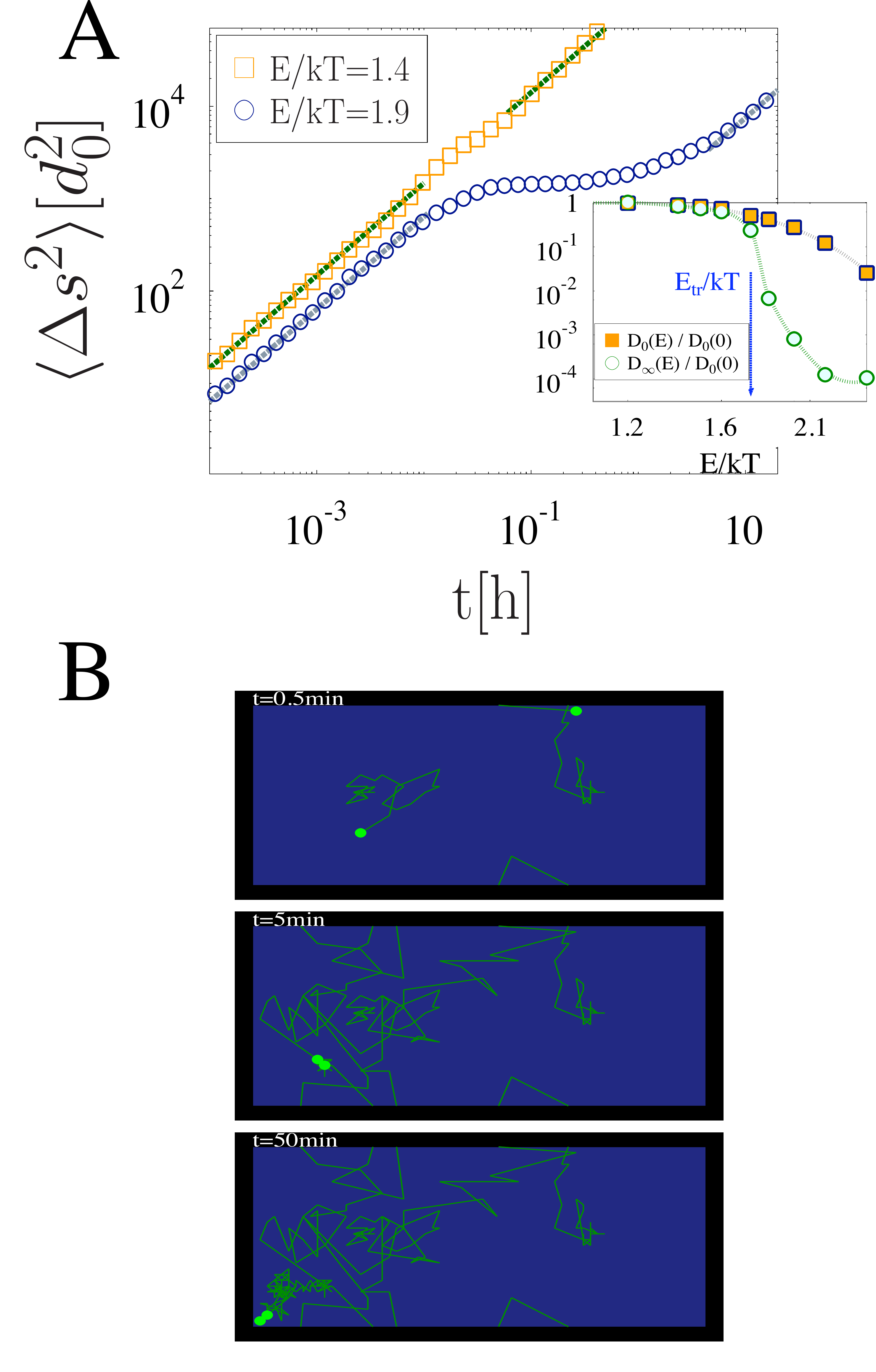}
\caption{
\textbf{Diffusion and colocalization of two DNA loci.} 
This figure illustrates the colocalization of two DNA segments. 
\textbf{Panel A} 
The mean square displacement of the center of mass of one of the DNA segments, $\langle \Delta s^{2}\rangle (t)$, is plotted as a function of time t. At small binding energies (squares), $\langle\Delta s^{2}\rangle (t)$ has a linear diffusive behaviour at all times as no colocalization occurs. At higher energies (circles), two different diffusive regimes are found at short and long time scales, before and after colocalization. In the inset, the diffusion constants at short and long time scales ($D_{0}$ and $D_{\infty}$) are shown as function of $E$. \textbf{Panel B} 2D projections of the system trajectory from a Monte Carlo simulation showing the initial Brownian diffusion of two separated DNA loci ($t=0.5$ minutes) and their colocalization ($t=5$ minutes and $t=50$ minutes).}\label{fig7}
\end{figure*}

Similar mechanisms act to shuttle a DNA segment to another DNA segment \cite{NicodemiPairing,ScialdonePairing}. We illustrate this by considering now a model that includes two polymers \cite{NicodemiPairing,ScialdonePairing}. As before, two regimes are found: when $E$, $c$ and $n_{0}$ are below threshold, the polymers float independently; above threshold, they colocalize.
When the colocalization machinery is switched on, the DNA segments will inevitably find and bind each other. Fig.~\ref{fig7}B illustrates different stages of the dynamics leading to colocalization: the polymer centers-of-mass are highlighted in green, whereas the darker green lines trace the trajectory they spanned up to that moment. The pictures show how, once their initial Brownian diffusion brings the DNA segments close enough (see $t=0.5$ minutes), i.e. within the range of the effective attraction induced by the MFs, they colocalize ($t=5$ minutes) and begin to diffuse together in the lattice ($t=50$ minutes).
The mean square displacement $\langle \Delta s^{2}\rangle(t)$ of the centers-of-mass of the DNA segments is plotted in Fig.~\ref{fig7}A. At low energy (e.g., $E/kT=1.4$), when the colocalization machinery is off, Brownian motion has approximately the same diffusion constant at short and long times. At higher energy (e.g., $E/kT=1.9$) two dynamical regimes are found: an initial one when the two polymers diffuse independently and a longer, slower diffusion when they move bound to each other. Such a behaviour is captured by a plot of the short- and long-time diffusion constants, $D_{0}$ and $D_{\infty}$, as function of $E$ (Fig.~\ref{fig7}A, inset). As shown earlier, $D_{0}(E)$ decreases with $E$, and $D_{\infty}(E)$ follows it. The transition point, $E_{tr}$, is marked by a drastic reduction of $D_{\infty}(E)$, whereas no major changes are found in the behaviour of $D_{0}(E)$. Above $E_{tr}$, $D_{\infty}(E)$ is non-zero as the two paired DNA segments continue to diffuse, although with a diffusion constant that is some orders of magnitude smaller than in the free case (Fig.~\ref{fig7}B). Such a large reduction is due to the much larger mass of the diffusing object in the colocalized state, which is formed by the couple of polymers and by a number of attached molecules.

\section*{Discussion}
\begin{figure*}
\begin{center}
\includegraphics[width=0.45\textwidth]{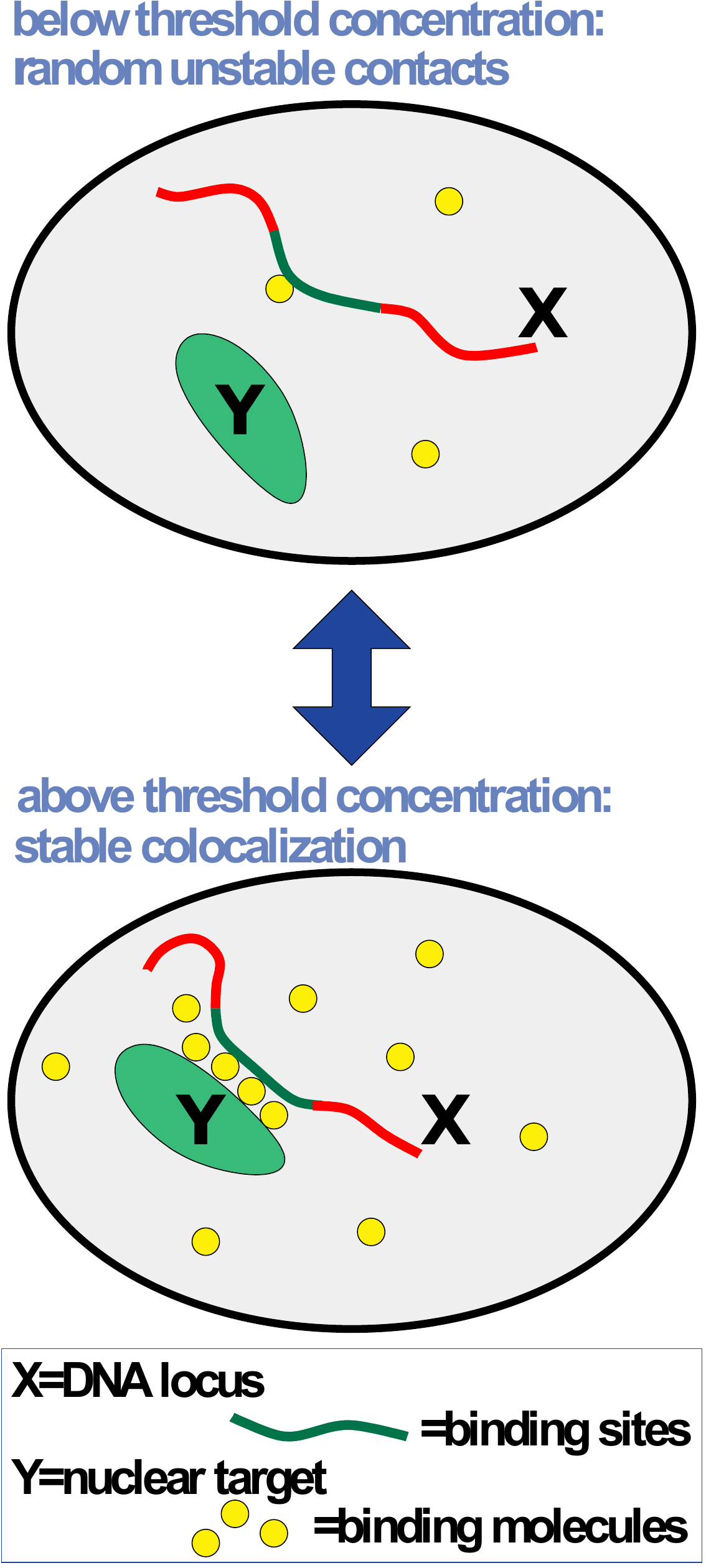}
\end{center}
\caption{
\textbf{Pictorial representation of the mechanism whereby molecular factors mediate DNA-target recognition and colocalization.} 
The DNA-target colocalization mechanism here investigated has a thermodynamic origin. It occurs as a switch-like process only when the concentration and the affinity of molecular binders exceed a threshold value corresponding to a phase transition (in a finite-sized system). Conversely, below the threshold, stable colocalization is thermodynamically impossible and the loci diffuse independently (see phase diagram in Figs.~\ref{fig1} and \ref{fig5}). The process has no energy costs, with resources being provided by thermal bath.
}\label{fig8}
\end{figure*}

In the cell nucleus, in a striking example of self-organization, the architecture of a vast number of DNA and nuclear loci is orchestrated to form complex and functional patterns involving regulatory cross-talks. In most cases, active processes are not required for colocalization \cite{deLaat2003,Dekker2008,Misteli2001} and questions arise on how DNA sequences recognize their targets and establish their relative positioning, and how the cell can control these processes.
Via a schematic statistical mechanics model, here we tried to address these questions and to propose a first quantitative scenario of a colocalization mechanism based on weak, biochemically unstable interactions between specific DNA sequences and their molecular binders. The mere production of molecules that bind both DNA and target is not sufficient to produce reliable and stable contacts. We showed they are activated only above a phase transition point, i.e. for concentration and affinity of the molecular mediators above precise threshold values (e.g., molecule concentrations around $\rho \sim10^{-3}-10^{-1} \mu mol/l$ correspond to transition energies in the range $E\sim 3-7kT$). Once these conditions are met, DNA loci find their relative positions as stable thermodynamic states at no energetic costs, as the resources required are provided by the surrounding thermal bath (Fig.~\ref{fig8}).
The switch-like nature of the mechanism of target recognition and colocalization we discussed could be exploited in the cell to reliably induce loci colocalization. In fact, well-known cell strategies of chromatin structure modification (i.e. change in $E$ or $n_{0}$) or upregulation of binding proteins (i.e. change in $c$) can produce precise, switch-like architectural rearrangements. Deep similarities are found across a variety of experimental data like those discussed in the Introduction, including specific aspects such as the effects of protein concentration changes on DNA looping (for examples, see \cite{Jing2008,Kumar2007}. The robust thermodynamic essence of the process we discuss could support the idea that passive shuttling phenomena can be traced back to simple universal mechanisms \cite{deGennes,deBell1993,Joanny1979,Semenov1996,Stanley}, in a sense independent of the biochemical details found in specific cases. Conversely, many complexities can arise in real cell nuclei, where a variety of other specific mechanisms are likely to intervene.
For sake of definiteness, we referred to DNA, but similar thermodynamic mechanisms could work for other biological polymers such as RNA, etc. Non-specific molecular factors and non-specific DNA binding can further assist the search kinetics \cite{BergO1981,Winter1981a,Winter1981b}, whereas other processes can intervene (e.g. to stabilize binding and to adjust DNA-target alignment if necessary). We also showed that non-specific binding sites on DNA and/or on its target can have an important effect on colocalization, yet the general scenario depicted above is unchanged. Testable predictions about the outcomes of, for example, genetic and/or chemical manipulations (such as DNA deletions), can be made, which can be tested against experimental data.
We tried to set the system parameters (e.g., the molecule concentration, the dynamics time scale) in a regime relevant to the real biological cases (see model description). Nevertheless, our model is very schematic and we included only the minimal molecular ingredients (i.e., molecular binders and specific DNA sites) that emerge from the experiments. However, a simple model could better serve the purpose to illustrate the core ingredients necessary for DNA target recognition (which can be traced back to polymer adsorption) and to depict a schematic, yet quantitative, scenario.

\bigskip

\textbf{Acknowledgments - } This work was supported by INFN, CINECA supercomputing project and CASPURĠs high performance computing services.


\end{document}